\documentclass[aps,preprint]{revtex4}%
\usepackage{amssymb}
\usepackage{amsmath}
\usepackage{amsfonts}
\usepackage{graphicx}%
\begin{document}
\title{Ward Identities and High Energy Scattering Amplitudes in String Theory }
\author{Chuan-Tsung Chan}
\email{ctchan@phys.cts.nthu.edu.tw}
\affiliation{Physics Division, National Center for Theoretical Sciences, Hsinchu, Taiwan, R.O.C.}
\author{Pei-Ming Ho}
\email{pmho@ntu.edu.tw}
\affiliation{Department of Physics, National Taiwan University and National Center for
Theoretical Sciences, Taipei, Taiwan, R.O.C.}
\author{Jen-Chi Lee}
\email{jcclee@cc.nctu.edu.tw}
\affiliation{Department of Electrophysics, National Chiao-Tung University, Hsinchu, Taiwan, R.O.C.}
\date{\today }

\begin{abstract}
High-energy limit $\alpha^{\prime}\rightarrow\infty$ of stringy Ward
identities derived from the decoupling of two types of zero-norm states in the
old covariant first quantized (OCFQ) spectrum of open bosonic string are used
to check the consistency of saddle point calculations of high energy
scattering amplitudes of Gross and Mende and Gross and Manes. Some
inconsistencies of their saddle point calculations are found even for the
string-tree scattering amplitudes of the \textit{excited} string states. We
discuss and calculate the missing terms of the calculation by those authors to
recover the stringy Ward identities. In addition, based on the tree-level
stringy Ward identities, we give the proof of a general formula, which was
proposed previously, of all high energy four-point string-tree amplitudes of
arbitrary particles in the string spectrum. In this formula all such
scattering amplitudes are expressed in terms of those of tachyons as
conjectured by Gross. The formula is extremely simple which manifestly
demonstrates the universal high energy behavior of the interactions among all
string states.

\end{abstract}
\maketitle


\section{\bigskip Introduction}

The study of high energy behavior of field theories, in particular Yang-Mills
theories, was very successful in the early 70's. In the quantum
chromodynamics, for example, the discovery of asymptotic freedom \cite{1}
turned out to be one of the most important properties of Yang-Mills theories.
On the other hand, the hidden spontaneously broken symmetry becomes evident at
high energies. It is thus very tempting to generalize this study to string
theory, which certainly contains a huge hidden symmetry. In 1988, Gross and
Mende \cite{2} proposed a saddle point method to calculate high energy
$\alpha^{\prime}\rightarrow\infty$ fixed angle string scattering amplitudes.
They identified a saddle point to the leading order in energy in the
calculation of first quantized string scattering amplitudes for all loops in
string perturbation theory. Soon after, based on this remarkable calculation,
Gross \cite{3} made important conjectures on high energy stringy symmetries.
There are two main conjectures of Gross's pioneer work on this subject. The
first one is the existence of an infinite number of linear relations among the
scattering amplitudes of different string states that are valid order by order
in perturbation theory at high energies. The second is that this symmetry is
so powerful as to determine the scattering amplitudes of all the infinite
number of string states in terms of, say, the dilaton (tachyon for the case of
open string) scattering amplitudes. However, the symmetry charges of his
proposed stringy symmetries were not understood and the proportionality
constants between scattering amplitudes of different string states were not
calculated. As we will see soon, all these problems can be solved by using
another independent calculation based on the following key idea: \textit{the
identification of symmetry charges from an infinite number of stringy
zero-norm states with arbitrarily high spins in the OCFQ spectrum}.

The importance of zero-norm states and their implication on stringy symmetries
were first pointed out in the context of the massive $\sigma$-model approach
of string theory \cite{4}. Zero-norm states were shown to imply
\textit{inter-particle symmetries} in the first order weak background field
approximation which is valid to all energies. On the other hand, zero-norm
states were also shown \cite{5} to carry the spacetime $\omega_{\infty}$
symmetry charges of the 2D string theory. Some implications of stringy Ward
identities, derived from the decoupling of two types of zero-norm states, on
stringy scattering amplitudes were also discussed in \cite{6}. Recently it was
discovered that \cite{7,8} the high energy limit of these stringy Ward
identities imply an infinite number of linear relations among scattering
amplitudes of different string states with the same momenta. These linear
relations can be used to fix the proportionality constants
\textit{algebraically }between scattering amplitudes of different string
states at each fixed mass level. These proportionality constants \ were found
to be independent of the scattering angle $\phi_{CM}$ and the loop order
$\chi$ of string perturbation theory as was conjectured by Gross \cite{3}.
Thus there is only one independent component of high energy stringy scattering
amplitudes for each fixed mass level. For the case of string-tree amplitudes,
a general formula can even be given to determine all high energy stringy
scattering amplitudes for arbitrary mass levels in terms of those of tachyons
\cite{7} - another conjecture by Gross \cite{3}.

Since the results we obtained from zero-norm states approach \cite{7,8} are
more complete than those obtained via the saddle point method in the past
\cite{2,3,11}, it would be of interest to directly compare these two
independent calculations. After a brief review of zero-norm state calculation
in section II, in section III we will use high energy limits of our stringy
Ward identities, which are valid to all energies and all loops, as a
consistent check of saddle point calculations. Since there is so far no
independent rigorous check of saddle point calculation, partially due to the
highly nontrivial structure of moduli spaces of general Riemann surfaces, our
simple stringy Ward identities serve as the best theoretical test of this
saddle point approximation. \textit{Surprisingly, some inconsistencies are
found for the results of saddle point calculations even for the string-tree
scattering amplitudes of the excited string states}. We discuss and calculate
the missing terms of the calculation by those authors to recover the stringy
Ward identities. In section IV of this paper, based on the stringy Ward
identities, we derive a general formula of all high energy four-point
string-tree amplitudes of arbitrary string states. This formula was first
proposed in \cite{7}. Here we will give a general proof and present some
examples. \textit{This formula determines the scattering amplitudes of all the
infinite number of string states in terms of tachyon scattering amplitudes}.
The formula is extremely simple which manifestly demonstrates the universal
high energy behavior of the interactions of an infinite number of string states.

\section{Zero-norm state calculation}

Let's begin with a brief review of zero-norm state calculation \cite{7,8}. In
the OCFQ spectrum of open bosonic string theory, the solutions of physical
state conditions include positive-norm propagating states and two types of
zero-norm states. The latter are \cite{9}%

\begin{equation}
\text{Type I}:L_{-1}\left\vert x\right\rangle ,\text{ where }L_{1}\left\vert
x\right\rangle =L_{2}\left\vert x\right\rangle =0,\text{ }L_{0}\left\vert
x\right\rangle =0; \tag{1}%
\end{equation}%
\begin{equation}
\text{Type II}:(L_{-2}+\frac{3}{2}L_{-1}^{2})\left\vert \widetilde
{x}\right\rangle ,\text{ where }L_{1}\left\vert \widetilde{x}\right\rangle
=L_{2}\left\vert \widetilde{x}\right\rangle =0,\text{ }(L_{0}+1)\left\vert
\widetilde{x}\right\rangle =0. \tag{2}%
\end{equation}
While type I states have zero-norm at any spacetime dimension, type II states
have zero-norm \textit{only} at D=26. In the first quantized approach of
string theory, the stringy \textit{on-shell} Ward identities are proposed to
be \cite{6}%
\begin{equation}
\mathcal{T}_{\chi}(k_{i})=g^{2-\chi}\int\frac{Dg_{\alpha\beta}}{\mathcal{N}%
}DX^{\mu}\exp(-\frac{\alpha^{\prime}}{2\pi}\int d^{2}\xi\sqrt{g}g^{\alpha
\beta}\partial_{\alpha}X^{\mu}\partial_{\beta}X_{\mu})\overset{4}%
{\underset{i=1}{\Pi}}v_{i}(k_{i})=0, \tag{3}%
\end{equation}
where at least one of the 4 vertex operators corresponds to the zero-norm
state. In eq.(3) $g$ is the string coupling constant, $\mathcal{N}$ is the
volume of the group of diffeomorphisms and Weyl rescalings of the worldsheet
metric, and $v_{i}(k_{i})$ are the on-shell vertex operators with momenta
$k_{i}$. The integral is over orientable open surfaces of Euler number $\chi$
parametrized by moduli $\overrightarrow{m}$ with punctures at $\xi_{i}$. In
this section, we will use stringy Ward identities of the second mass level
\ $m^{2}$ $=4$ as an example to illustrate our approach. The four stringy Ward
identities at this mass level were calculated to be \cite{6}%
\begin{equation}
k_{\mu}\theta_{\nu\lambda}\mathcal{T}_{\chi}^{(\mu\nu\lambda)}+2\theta_{\mu
\nu}\mathcal{T}_{\chi}^{(\mu\nu)}=0, \tag{4}%
\end{equation}%
\begin{equation}
(\frac{5}{2}k_{\mu}k_{\nu}\theta_{\lambda}^{\prime}+\eta_{\mu\nu}%
\theta_{\lambda}^{\prime})\mathcal{T}_{\chi}^{(\mu\nu\lambda)}+9k_{\mu}%
\theta_{\nu}^{\prime}\mathcal{T}_{\chi}^{(\mu\nu)}+6\theta_{\mu}^{\prime
}\mathcal{T}_{\chi}^{\mu}=0, \tag{5}%
\end{equation}%
\begin{equation}
(\frac{1}{2}k_{\mu}k_{\nu}\theta_{\lambda}+2\eta_{\mu\nu}\theta_{\lambda
})\mathcal{T}_{\chi}^{(\mu\nu\lambda)}+9k_{\mu}\theta_{\nu}\mathcal{T}_{\chi
}^{[\mu\nu]}-6\theta_{\mu}\mathcal{T}_{\chi}^{\mu}=0, \tag{6}%
\end{equation}%
\begin{equation}
(\frac{17}{4}k_{\mu}k_{\nu}k_{\lambda}+\frac{9}{2}\eta_{\mu\nu}k_{\lambda
})\mathcal{T}_{\chi}^{(\mu\nu\lambda)}+(9\eta_{\mu\nu}+21k_{\mu}k_{\nu
})\mathcal{T}_{\chi}^{(\mu\nu)}+25k_{\mu}\mathcal{T}_{\chi}^{\mu}=0, \tag{7}%
\end{equation}
where $\theta_{\mu\nu}$ is transverse and traceless, and $\theta_{\lambda
}^{\prime}$ and $\theta_{\lambda}$ are transverse vectors. These are
polarizations of zero-norm states. In each equation, we have chosen, say,
$v_{2}(k_{2})$\ to be the vertex operators constructed from zero-norm states
at the mass level $m^{2}$ $=4$ and $k_{\mu}\equiv k_{2\mu}$. In eqs. (4)-(7),
$v_{1},$ $v_{3}$ and $v_{4}$ can be any string states (including zero-norm
states) at any mass level and we have omitted their tensor indices.
$\mathcal{T}_{\chi}^{\prime}s$ in eqs (4)-(7) are $\chi$-th order string-loop
amplitudes. It is important to note that eqs. (4)-(7) are valid to all loop
orders $\chi$ and all energies $E$, and are \emph{automatically} of the
identical form in string perturbation theory.

We will use labels 1 and 2 for the incoming particles and 3 and 4 for the
outgoing particles. The center of mass scattering angle $\phi_{CM}$ is defined
to be the angle between $\overrightarrow{k}_{1}$ and $\overrightarrow{k}_{3}$.
The leading order Ward identities of eqs.(4) -(7) in the energy $E$ expansions
were calculated to be (we drop loop order $\chi$ here to simplify the
notation)\cite{7,8}%
\begin{equation}
\mathcal{T}_{LLT}^{5\rightarrow3}+\mathcal{T}_{(LT)}^{3}=0, \tag{8}%
\end{equation}%
\begin{equation}
10\mathcal{T}_{LLT}^{5\rightarrow3}+\mathcal{T}_{TTT}^{3}+18\mathcal{T}%
_{(LT)}^{3}=0, \tag{9}%
\end{equation}%
\begin{equation}
\mathcal{T}_{LLT}^{5\rightarrow3}+\mathcal{T}_{TTT}^{3}+9\mathcal{T}%
_{[LT]}^{3}=0, \tag{10}%
\end{equation}
where the subscripts and superscripts denote the polarizations and energy
orders respectively, which will be explained below. A simple calculation shows
that%
\begin{equation}
\mathcal{T}_{TTT}^{3}:\mathcal{T}_{LLT}^{3}:\mathcal{T}_{(LT)}^{3}%
:\mathcal{T}_{[LT]}^{3}=8:1:-1:-1. \tag{11}%
\end{equation}
In the above equations, we have defined the normalized polarization vectors of
the second string state to be%
\begin{equation}
e_{P}=\frac{1}{m_{2}}(E_{2},\mathrm{k}_{2},0)=\frac{k_{2}}{m_{2}}, \tag{12}%
\end{equation}%
\begin{equation}
e_{L}=\frac{1}{m_{2}}(\mathrm{k}_{2},E_{2},0), \tag{13}%
\end{equation}%
\begin{equation}
e_{T}=(0,0,1) \tag{14}%
\end{equation}
in the CM frame contained in the plane of scattering. $\mathcal{T}_{TTT}%
=e_{T}^{\mu}e_{T}^{\nu}e_{T}^{\lambda}$ $\mathcal{T}_{\mu\nu\lambda}$ etc. In
eqs.(8) - (10), we have assigned a relative energy power to each amplitude.
For each longitudinal $L$ component, the order is $E^{2}$ ( the naive order of
$e_{L}\cdot k$ is $E^{2}$ ) and for each transverse $T$ component, the order
is $E$ ( the naive order of $e_{T}\cdot k$ is $E$ )$.$ This is due to the
definitions of $\ e_{L}$ and $e_{T}$ in eqs (13) and (14), where $e_{L}$ got
one energy power more than $e_{T}.$ Due to eq.(8), the \textit{naive leading
order} $E^{5}$ term of the energy expansion for $\mathcal{T}_{LLT}$ is forced
to be zero. As a result, the true leading order is at most $E^{3}$. This is
the meaning of the superscript $5\rightarrow3$ in eq.(8). Similar rule applies
to other equations. It is important to note that eqs.(8) - (11) are valid to
all loops and are independent of the particles chosen for $v_{1,3,4}$. For the
case of string-tree level $\chi=1$ with one tensor $v_{2}$ and three tachyons
$v_{1,3,4}$, all four scattering amplitudes in eq.(11) were calculated to be
$\mathcal{T}_{TTT}^{3}=-8E^{9}\sin^{3}\phi_{CM}\mathcal{T}(3)=8\mathcal{T}%
_{LLT}^{3}=-8\mathcal{T}_{(LT)}^{3}=-8\mathcal{T}_{[LT]}^{3}$ , where%
\begin{align}
\mathcal{T}(n)  &  =\sqrt{\pi}(-1)^{n-1}2^{-n}E^{-1-2n}(\sin\frac{\phi_{CM}%
}{2})^{-3}(\cos\frac{\phi_{CM}}{2})^{5-2n}\nonumber\\
&  \times\exp(-\frac{s\ln s+t\ln t-(s+t)\ln(s+t)}{2}), \tag{15}%
\end{align}
is the high energy limit of $\frac{\Gamma(-\frac{s}{2}-1)\Gamma(-\frac{t}%
{2}-1)}{\Gamma(\frac{u}{2}+2)}$ with $s+t+u=2n-8$, and we have calculated it
up to the next leading order in $E$. This is the order which includes the
energy power factor in front of the exponential. Eq.(11) was thus explicitly
justified \cite{7,8}. In eq.(15) $%
s=-(k_{1}+k_{2})^{2}%
$, $%
t=-(k_{2}+k_{3})^{2}%
$ and $%
u=-(k_{1}+k_{3})^{2}%
$ are the Mandelstam variables and our convention here is different from
references \cite{2,3,11} by interchanging $t\longleftrightarrow u$.

In deriving eqs. (8) - (10), we have identified $\mathcal{T}_{\cdot\cdot
P\cdot\cdot}=\mathcal{T}_{\cdot\cdot L\cdot\cdot}$ not only at the naive
leading order but also at the true leading order in energy. For the massless
case, this is true since by definitions eqs.(12) and (13), $e_{P}=e_{L}$.
However, for the massive case, it is not obvious that they can be identified.
Naively, in the high energy limit, all masses go to zero, and one expects
smooth massless limits for all relevant physical quantities and $\mathcal{T}%
_{\cdot\cdot P\cdot\cdot}=\mathcal{T}_{\cdot\cdot L\cdot\cdot}$ as in the
massless case. This issue is a more familiar subject in the context of field
theories and it turns out that the smooth massless limit may not be achieved
for an arbitrary massive field theory. For example, it is well known that the
massless limit of a massive gauge field theory is in general different from
the massless theory. To illustrate this point, consider a massive gauge field
with the Lagrangian density
\begin{equation}
\mathcal{L}=\frac{1}{4}F_{\mu\nu}F^{\mu\nu}+\frac{m^{2}}{2}A_{\mu}A^{\mu
}+A_{\mu}J^{\mu}+\cdot\cdot, \tag{16}%
\end{equation}
where $J$ is a current coupled to $A$ and $\cdot\cdot$ represents the kinetic
term and possible interaction terms for $J$. The equation of motion of $A$ is
solved in momentum space as
\begin{equation}
A_{\mu}=\frac{1}{k^{2}+m^{2}}\left(  \eta_{\mu\nu}+\frac{1}{m^{2}}k_{\mu
}k_{\nu}\right)  J^{\nu}. \tag{17}%
\end{equation}
Immediately we see that the massless limit is discontinuous since the second
term on the right blows up when $m^{2}\rightarrow0$. Fronsdal \cite{10} showed
that, for vector gauge potentials as well as gauge fields of higher spins, the
requirement of the continuity of the massless limit is equivalent to the
conservation of charge associated with the gauge symmetry. The second term in
eq.(17) vanishes if we assume charge conservation $k_{\mu}J^{\mu}=0$. Note
that the assumption of charge conservation also leads to Ward identities
$k_{\mu}\langle J^{\mu}\cdots\rangle=0$, which are certainly nontrivial
relations among correlation functions.

For the cases of our stringy massive states, we do have these stringy gauge
symmetries or conserved charges to fulfil Fronsdal's criterion. Therefore,
although naively zero-norm states can not by themselves establish nontrivial
relations among scattering amplitudes, our assumption about the continuity of
the high energy limit, which is implicitly imposed when we identify
$\mathcal{T}_{\cdot\cdot P\cdot\cdot}=\mathcal{T}_{\cdot\cdot L\cdot\cdot}$,
leads to nontrivial relations, namely our stringy Ward identities. In our
prescription, zero-norm state is a vehicle used to bring the information about
charge conservation to the surface to be seen. As we found explicitly for mass
levels $m^{2}=4$ and $m^{2}=6$ \cite{7,8}, our prescription indeed leads to
ratios between scattering amplitudes for different particles. We give one
example here to justify our assumption $\mathcal{T}_{\cdot\cdot P\cdot\cdot
}=\mathcal{T}_{\cdot\cdot L\cdot\cdot}$. We have explicitly checked that%
\begin{equation}
\mathcal{T}_{PPT}^{3}=\mathcal{T}_{LPT}^{3}=\mathcal{T}_{LLT}^{3}=-E^{9}%
\sin^{3}\phi_{CM}\mathcal{T}(3), \tag{18}%
\end{equation}
to the leading order in energy, for the case of string tree amplitudes with
spin-three tensor $v_{2}$ and three tachyons $v_{1,3,4}$.

\section{Saddle point calculation corrected}

We now briefly review the saddle point calculation of eq.(3) \cite{2,3,11}.
First one notes that the high energy limit $\alpha^{\prime}\rightarrow\infty$
is equivalent to the semi-classical limit of first-quantized string theory. In
this limit, the closed string $G$-loop scattering amplitudes is dominated by a
saddle point in the moduli space $\overrightarrow{m}$. For the oriented open
string amplitudes, the saddle point configuration can be constructed from an
associated configuration of the closed string via reflection principle. It was
also found that the Euler number $\chi$ of the oriented open string saddle is
always $%
\chi=1-G%
$, where $G$ is the genus of the associated closed string saddle. Thus the
integral in eq. (3) is dominated in the $\alpha^{\prime}\rightarrow\infty$
limit by an associated $G$-loop closed string saddle point in $X^{\mu}%
$,$\widehat{\overrightarrow{m}_{i}}$ and $\widehat{\xi_{i}}$. The closed
string classical trajectory at $G$-loop order was found, according to Gross
and Mende \cite{2}, to behave at the saddle point as%
\begin{equation}
X_{c1}^{\mu}(z)=\frac{i}{1+G}\overset{4}{\underset{i=1}{\sum}}k_{i}^{\mu}%
\ln\left\vert z-a_{i}\right\vert +O(\frac{1}{\alpha^{\prime}}), \tag{19}%
\end{equation}
which leads to the $\chi$-th order open string four-tachyon amplitude%
\begin{equation}
\mathcal{T}_{\chi}\approx g^{2-\chi}\exp(-\frac{s\ln s+t\ln t+u\ln u}%
{2(2-\chi)}). \tag{20}%
\end{equation}
Eq.(20) reproduces the very soft exponential decay e$^{-\alpha^{\prime}s}$ of
the well-known string-tree $\chi$=1 amplitude. There is a consistent check of
eq.(20) at small angles, where the genus-$G$ scattering process can be
decomposed into $G+1$ successive scatterings \cite{2,12}. The exponent of
eq.(20) can be thought of as the electrostatic energy $E_{G}$ of
two-dimensional Minkowski charges $k_{i}$ placed at $a_{i}$ on a Riemann
surface of genus $G$. One can use the $SL(2,C)$ invariance of the saddle to
fix 3 of the 4 points $a_{i}$, then the only modulus is the cross ratio
$\lambda=\frac{(a_{1}-a_{3})(a_{2}-a_{4})}{(a_{1}-a_{2})(a_{3}-a_{4})}$, which
takes the value%
\begin{equation}
\lambda=\widehat{\lambda}\approx-\frac{t}{s}\approx\sin^{2}\frac{\phi_{CM}}%
{2}\text{\ \ \ \ \ \ \ \ \ \ \ \ \ \ \ \ \ \ \ \ \ \ \ \ \ \ \ } \tag{21}%
\end{equation}
to extremize $E_{G}$ if we neglect the mass of the tachyons in the high energy
limit. For excited string states, it was found that only polarizations in the
plane of scattering will contribute to the amplitude at high energies. To the
leading order in energy $E$, the products of e$_{T}$ and e$_{L}$ with
$\partial^{n}X$ at string-tree level $\chi=1$ (or $G=0$) were calculated by
using eq.(19) to be \cite{11}%
\begin{equation}
e_{T}\cdot\partial^{n}X\sim i(-)^{n}\frac{(n-1)!}{\lambda^{n}}E\sin\phi
_{CM},\text{ }n>0; \tag{22}%
\end{equation}%
\begin{equation}
e_{L}\cdot\partial^{n}X\sim i(-)^{(n-1)}\frac{(n-1)!}{\lambda^{n}}\frac
{E^{2}\sin^{2}\phi_{CM}}{2m_{2}}\overset{n-2}{\underset{l=0}{\sum}\lambda^{l}%
},\text{ }n>1; \tag{23}%
\end{equation}%
\begin{equation}
e_{L}\cdot\partial^{n}X\sim0,\text{ }%
n=1,\text{\ \ \ \ \ \ \ \ \ \ \ \ \ \ \ \ \ \ \ \ \ \ \ \ \ \ \ \ \ } \tag{24}%
\end{equation}
where $m_{2}$ is the mass of the particle.

\bigskip We are now ready to use the results of our zero-norm state
calculation, eqs.(8) - (11), to check the validity of the saddle point
calculations eqs.(21) - (24). Note that eqs.(8) - (11) are the high energy
limit of the decoupling of zero-norm states of stringy scattering amplitudes.
They are directly related to the unitarity of string theory. We will just
check the string-tree amplitudes since only in this case the exact results are
known. Let's use eqs.(21)-(24) to calculate, for example, the high energy
limit of eq.(4) where both scattering amplitudes $\mathcal{T}^{(\mu\nu
\lambda)},$ $\mathcal{T}^{(\mu\nu)}$ are defined by eq.(3). It is easy to see,
according to eqs.(21) - (24), that the kinematic factor $\mathcal{K}_{LLT}$ of
$\mathcal{T}_{LLT}$ is of energy order $E$ while $\mathcal{K}_{(LT)}\sim
E^{3}$. This means that $\mathcal{T}_{LLT}^{3}=0$ \cite{7,8} to the leading
order in the saddle point calculation. This obviously violates the result of
our zero-norm state calculation eq.(8), which says $\mathcal{K}_{LLT}^{3}$ and
$\mathcal{K}_{(LT)}^{3}$ are of the same energy order $E^{3}$. Since eq.(8) is
the high energy limit of eq.(4). We conclude that there is an inconsistency
between eq.(4) and eq.(21)-(24). In other words,\textit{\ the results
eqs.(21)-(24) violates the high energy massive gauge invariance \cite{7,8} of
eq.(8) and thus will threat the unitarity of string interactions.} Many
similar examples with the same inconsistencies exist at higher mass levels
mainly due to the wrong result of eq.(24). They are scattering processes with
vertices containing $\partial X^{L}$, or processes whose naive leading order
in energy vanishes, e.g.$\mathcal{T}_{LLT}^{5\rightarrow3}$.

To further demonstrate why the previous saddle point calculations
\cite{2,3,11} fail to be consistent with the stringy Ward identity, we redo
the calculations based on the saddle point method and figure out the missing
terms in the previous results. Again, we shall use the examples of $m^{2}=4$
amplitudes $\mathcal{T}_{LLT}$ and $\mathcal{T}_{TTT}$ to demonstrate the
correct calculation. The spin-three amplitude is defined as
\begin{equation}
\mathcal{T}^{\mu\nu\lambda}\equiv%
{\textstyle\int}
{\textstyle\prod_{i=1}^{4}}
dx_{i}<e^{ik_{1}X}\partial X^{\mu}\partial X^{\nu}\partial X^{\lambda
}e^{ik_{2}X}e^{ik_{3}X}e^{ik_{4}X}>, \tag{25}%
\end{equation}
and $\mathcal{T}_{LLT}=e_{L}^{\mu}e_{L}^{\nu}e_{T}^{\lambda}\mathcal{T}%
_{\mu\nu\lambda}$ is calculated to be
\begin{align}
\mathcal{T}_{LLT}  &  =\int_{0}^{1}dx\hspace{0.3cm}x^{-s/2-1}(1-x)^{-t/2}%
(e_{L}\cdot k_{1})(e_{L}\cdot k_{1})(e_{T}\cdot k_{3})\tag{26}\\
&  -2\int_{0}^{1}dx\hspace{0.3cm}x^{-s/2}(1-x)^{-t/2-1}(e_{L}\cdot
k_{1})(e_{L}\cdot k_{3})(e_{T}\cdot k_{3})\tag{27}\\
&  +\int_{0}^{1}dx\hspace{0.3cm}x^{-s/2+1}(1-x)^{-t/2-2}(e_{L}\cdot
k_{3})(e_{L}\cdot k_{3})(e_{T}\cdot k_{3}). \tag{28}%
\end{align}
\bigskip Similarly, for $\mathcal{T}_{TTT}$, we have
\begin{equation}
\mathcal{T}_{TTT}=\int_{0}^{1}dx\hspace{0.3cm}x^{-s/2+1}(1-x)^{-t/2-2}%
(e_{T}\cdot k_{3})(e_{T}\cdot k_{3})(e_{T}\cdot k_{3}). \tag{29}%
\end{equation}
In deriving eq.(26)-(29), we have made the $SL(2,R)$ gauge fixing and
restricted to the $s-t$ channel amplitude only by choosing $x_{1}=0,0\leqq
x_{2}\leqq1,x_{3}=1,x_{4}=\infty.$ Here we need to evaluate three integrals by
using saddle point method. For instance, the integral of eqs.(28) and (29) can
be defined as
\begin{align}
F_{-1,2}\equiv\int_{0}^{1}dx\hspace{0.3cm}x^{-s/2+1}(1-x)^{-t/2-2}  &
=\int_{0}^{1}dx\hspace{0.3cm}x(1-x)^{-2}\exp^{-\frac{s}{2}[\ln x-\tau
\ln(1-x)]}\nonumber\\
&  =\int_{0}^{1}dx\hspace{0.3cm}u_{-1,2}(x)\exp^{-\frac{s}{2}f(x)}, \tag{30}%
\end{align}
where we have defined
\begin{align}
\tau &  \equiv-\frac{t}{s}\simeq\sin^{2}\frac{\phi_{CM}}{2},\tag{31}\\
u_{a,b}(x)  &  \equiv x^{-a}(1-x)^{-b},\tag{32}\\
f(x)  &  \equiv\ln x-\tau\ln(1-x). \tag{33}%
\end{align}
The other integrals in eqs.(26) and (27), $F_{1,0}$ and $F_{0,1},$ can be
similarily defined. Note that our definitions of Mandelstam variables here are
different from eqs. (21)- (24) by interchanging $t\longleftrightarrow u.$ Now,
before doing saddle point calculations, we would like to point out that there
are two different definitions of the concepts of saddle-points. For the
integral of eq.(30), for example, the first definition is the saddle point
$x_{0}$ which leads to $f^{\prime}(x)=0$ and is given by $x_{0}=\frac
{1}{1-\tau}$, and we have $f^{\prime\prime}(x_{0})=\frac{(1-\tau)^{3}}{\tau}$.
In this case, the saddle point is independent of the prefactor $u_{-1,2}.$ The
more standard definition of saddle point is, however, the value $x_{0}%
^{\prime}$ which extremizes the exponent of $\exp^{-\frac{s}{2}f(x)+\ln
u_{-1,2}}.$ In this case, the saddle points of the integrals of the scattering
amplitudes of the excited string states will be shifted level by level. We
stress that although this shift is of subleading order in energy compared with
eq.(20), its effect will bring down an energy power factor in front of the
exponential. These power factors are crucial to recover the stringy Ward
identities and get the linear relations among high energy scattering
amplitudes conjectured by Gross. In the following, we will adopt the first
definition to do the calculation. However, both saddle point calculations
should give the same results if one does the calculations carefully.

We can use the following formula for a systematic expansion of intergal in
terms of $\alpha^{\prime}$,
\begin{align}
F(\alpha^{\prime})  &  \equiv\int_{-\infty}^{\infty}dx\hspace{0.3cm}%
u(x)\exp^{-\alpha^{\prime}sf(x)}\tag{34}\\
&  =u_{0}\exp^{-\alpha^{\prime}sf_{0}}\sqrt{\frac{2\pi}{\alpha^{\prime}%
sf_{0}^{\prime\prime}}}\nonumber\\
&  \times\{1+[\frac{u_{0}^{\prime\prime}}{2u_{0}f_{0}^{\prime\prime}}%
-\frac{u_{0}^{\prime}f_{0}^{(3)}}{2u_{0}(f_{0}^{\prime\prime})^{2}}%
-\frac{f_{0}^{(4)}}{8(f_{0}^{\prime\prime})^{2}}+\frac{5[f_{0}^{(3)}]^{2}%
}{24(f_{0}^{\prime\prime})^{3}}]\frac{1}{\alpha^{\prime}s}+O(\frac{1}%
{({\alpha^{\prime}s)}^{2}})\}, \tag{35}%
\end{align}
where $u_{0},f_{0},u_{0}^{\prime},f_{0}^{\prime\prime},$ etc, stand for the
values of functions and their derivatives evaluated at $x_{0}$. For
simplicity, we only write down the leading and next-to-leading corrections in
$\frac{1}{\alpha^{\prime}}$. One can extend this formula to higher orders such
that the desired accuracy can be achieved. Note that the range of integration
in eqs.(26)-(30) can be extended from $(0,1)$ to $(-\infty,+\infty)$ by a
change of variable $x\rightarrow\frac{e^{x}}{e^{x}+e^{-x}}.$

Now, in order to avoid complicated expansion of momentum variables $e_{L}\cdot
k_{1}$ and $e_{L}\cdot k_{3}$ in the calculations of $\mathcal{T}_{LLT}$
\cite{8}, we can use the previous result of $\mathcal{T}_{LLT}=\mathcal{T}%
_{PPT}$ to the leading order, and calculate
\begin{align}
\mathcal{T}_{LLT}\simeq\mathcal{T}_{PPT} &  =(\frac{E}{4}\sin\phi_{CM})\left[
F_{1,0}(k_{1}\cdot k_{2})^{2}-2F_{0,1}(k_{1}\cdot k_{2})(k_{2}\cdot
k_{3})+F_{-1,2}(k_{2}\cdot k_{3})^{2}\right]  \nonumber\\
&  =(E^{5}\sin\phi_{CM})\left[  F_{1,0}+2\tau F_{0,1}+\tau^{2}F_{-1,2}\right]
,\tag{36}\\
\mathcal{T}_{TTT} &  =(E\sin\phi_{CM})^{3}F_{-1,2}.\tag{37}%
\end{align}
The remaining task for calculating both amplitudes now amounts to the various
contributions of the master formula for $F_{1,0},F_{0,1}$ and $F_{-1,2}$. For
$\mathcal{T}_{LLT}$, we notice that $(u_{1,0}+2\tau u_{0,1}+\tau^{2}%
u_{-1,2})\mid_{x_{0}}=0$, hence both the leading terms and the last two terms
in the $\frac{1}{\alpha^{\prime}}$ corrections of eq.(35) cancel. After some
calculations, we find $(u_{1,0}^{\prime}+2\tau u_{0,1}^{\prime}+\tau
^{2}u_{-1,2}^{\prime})\mid_{x_{0}}=0$, so the second term in the $\frac
{1}{\alpha^{\prime}}$ corrections of eq.(35) sum up to zero. The only
surviving contribution comes from the first term in the $\frac{1}%
{\alpha^{\prime}}$ corrections in eq.(35), and we have
\begin{equation}
\frac{1}{2f_{0}^{\prime\prime}}\left[  u_{1,0}^{\prime\prime}+2\tau
u_{0,1}^{\prime\prime}+\tau^{2}u_{-1,2}^{\prime\prime}\right]  _{x_{0}}%
=\frac{(1-\tau)^{2}}{\tau}.\tag{38}%
\end{equation}
We emphasize that these are exactly the missing terms of the previous saddle
point calculations \cite{2,3,11}. Once these corrections are taken into
account \cite{13}, we are able to obtain the result which is consistent with
our stringy Ward identities. In fact, we can now use $\sin\phi_{CM}%
\simeq2\sqrt{\tau(1-\tau)}$ , $\sqrt{\frac{4\pi}{sf_{0}^{\prime\prime}}}%
=\frac{1}{E}\sqrt{\frac{\pi\tau}{(1-\tau)^{3}}}$ and eq.(38) to get (
$\alpha^{\prime}\equiv\frac{1}{2},$ $s\rightarrow\infty$ )
\begin{align}
\mathcal{T}_{LLT} &  =-(E^{5}\sin\phi_{CM})(\exp^{-\frac{s}{2}f_{0}}%
\sqrt{\frac{4\pi}{sf_{0}^{\prime\prime}}})\frac{(1-\tau)^{2}}{\tau}(\frac
{2}{s})=-\sqrt{\pi}E^{2}\cos^{2}\frac{\phi_{CM}}{2}\exp^{-\frac{s}{2}f_{0}%
}\nonumber\\
&  =-E^{9}\sin^{3}\phi_{CM}\mathcal{T}(3).\tag{39}%
\end{align}
Similarly, for $\mathcal{T}_{TTT}$, we have only one leading term in eq.(35),
and $u_{-1,2}\mid_{x_{0}}=\frac{1-\tau}{\tau^{2}}$ gives
\begin{align}
\mathcal{T}_{TTT} &  =-(E^{3}\sin^{3}\phi_{CM})(\exp^{-\frac{s}{2}f_{0}}%
\sqrt{\frac{4\pi}{sf_{0}^{\prime\prime}}})\frac{(1-\tau)}{\tau^{2}}%
=-8\sqrt{\pi}E^{2}\cos^{2}\frac{\phi_{CM}}{2}\exp^{-\frac{s}{2}f_{0}%
}\nonumber\\
&  =-8E^{9}\sin^{3}\phi_{CM}\mathcal{T}(3).\tag{40}%
\end{align}
Eqs.(39) and (40) agree with our previous calculations above eq.(15) based on
a different method \cite{7,8}. Finally, by comparing eq.(39) and eq.(40), we
obtain the desired relation, $\mathcal{T}_{TTT}:\mathcal{T}_{LLT}=8:1$.

In conclusion, we see that the use of saddle point in eqs.(19) and (21) is
only valid for the tachyons amplitude in eq.(20). \textit{In general, the
prediction of eq.(21)-(24) gives the right energy exponent in the scattering
amplitudes, but not the energy power factors in front of the exponential for
the cases of the excited string states.} \textit{These energy power factors
are subleading terms ignored in eqs.(21)-(24) but they are crucial if one
wants to get the linear relations among high energy scattering amplitudes
conjectured by Gross \cite{3}.}

\section{\bigskip String-tree high Energy scattering amplitudes}

In this section, we will first give a general formula of all high energy
four-point string-tree amplitudes of \textit{arbitrary string states}. This
formula was first proposed in \cite{7}. Here we will give a general proof and
present some examples. Let's begin with the scattering amplitudes of one
arbitrary tensor $v_{2}$ and three tachyons $v_{1,3,4}$. Based on the result
of high energy stringy Ward identities, it was conjectured in \cite{7,8} that
there is only one independent component of high energy scattering amplitude at
each fixed mass level, say, $\mathcal{T}_{n}^{TTT\cdot\cdot}$ which is defined
to be the transverse component of scattering amplitude of the highest spin
state at the $m^{2}$ $=2(n-1)$ level, where $n$ is the number of $T^{\prime}%
s$. For example, the first scattering in eq.(11) $\mathcal{T}_{3}^{TTT}$
corresponds to $m^{2}$ $=4$. All other components of high energy scattering
amplitudes are proportional to it. This conjecture was explicitly proved for
the mass level $m^{2}=4,6$. It is not difficult to calculate the following
general scattering amplitudes of this type with $v_{2}$ the highest spin state
at each fixed mass level and three tachyons $v_{1,3,4}$(we list amplitudes for
the $s-t$ channel only)%
\begin{equation}
\mathcal{T}_{n}^{\mu_{1}\mu_{2}\cdot\cdot\mu_{n}}=\overset{n}{\underset
{l=0}{\sum}}(-)^{l}\left(  _{l}^{n}\right)  B(-\frac{s}{2}-1+l,-\frac{t}%
{2}+n-l)k_{1}^{(\mu_{1}}..k_{1}^{\mu_{n-l}}k_{3}^{\mu_{n-l+1}}..k_{3}^{\mu
_{n})}, \tag{41}%
\end{equation}
where $B(u,v)=\int_{0}^{1}dxx^{u-1}(1-x)^{v-1}$ is the Euler beta function. It
is now easy to calculate the general high energy scattering amplitude at the
$m^{2}$ $=2(n-1)$ level \cite{8}%
\begin{equation}
\mathcal{T}_{n}^{TTT\cdot\cdot}=[-2E^{3}\sin\phi_{CM}]^{n}\mathcal{T}(n),
\tag{42}%
\end{equation}
where $\mathcal{T}(n)$ is given by eq.(15). One can now generalize this result
to multi-tensors \cite{7}%
\begin{equation}
\mathcal{T}_{n_{1}n_{2}n_{3}n_{4}}^{T^{1}\cdot\cdot T^{2}\cdot\cdot T^{3}%
\cdot\cdot T^{4}\cdot\cdot}=[-2E^{3}\sin\phi_{CM}]^{\Sigma n_{i}}%
\mathcal{T}(n\rightarrow\Sigma n_{i}), \tag{43}%
\end{equation}
\begin{figure}[th]
\includegraphics[width=7.5cm]{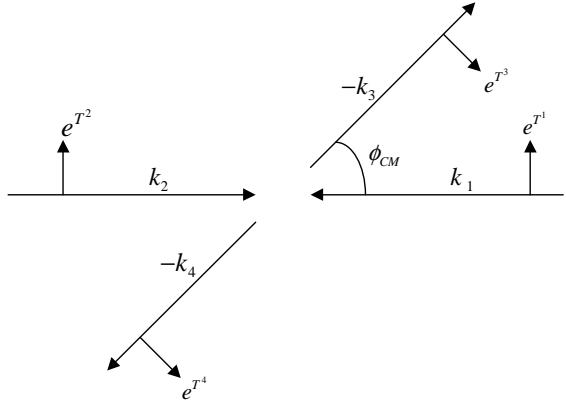}\caption{Kinematic variables in the
center of mass frame.}%
\end{figure}where $n_{i}$ is the number of $T^{i}$ of the $i-th$ vertex
operators and $T^{i}$ is the transverse direction of the $i-th$ particle. This
is our first master formula for all scattering amplitudes which are of the
leading order in energy compared with other amplitudes at the same mass
levels. In proving eq.(43), one notes that the new contraction terms arising
from two $\partial X^{\prime}s$ belonging to two different vertex operators
are suppressed in energy. Finally, the $t-u$ and $s-u$ channels of a given
scattering amplitude can be Mobius transformed to the $s-t$ channel of another
scattering amplitude obtained by interchanging vertex operators of the
original scattering amplitude. Thus we still end up with the $\sin\phi_{CM}$
behavior in the kinematic factor. This completes our proof for eq.(43). It is
remarkable to find, from eq.(43), that \textit{all} high energy scattering
amplitudes share this simple high energy behavior, which hints at the
universal form of the interactions of an infinite number of string states. In
the following we give some examples to illustrate the power of eq.(43). The
scattering amplitudes involving tachyons and massless vectors were calculated
in \cite{14}, and are given by ($s-t$ channel only)%
\begin{equation}
\mathcal{T}_{vectors}\mathcal{=}\frac{\Gamma(-\frac{s}{2}-1)\Gamma(-\frac
{t}{2}-1)}{\Gamma(\frac{u}{2}+2)}K, \tag{44}%
\end{equation}
with the kinematic factors,%
\begin{align}
K(\zeta_{1},k_{1};\zeta_{2},k_{2};k_{3};k_{4})  &  =(\frac{1}{2}t+1)(\frac
{1}{2}u+1)(\zeta_{1}\cdot k_{2}\zeta_{2}\cdot k_{1}-\zeta_{1}\cdot\zeta
_{2})\tag{45}\\
&  +(\frac{1}{2}s+1)(\frac{1}{2}u+1)(\zeta_{1}\cdot k_{4}\zeta_{2}\cdot
k_{3})\nonumber\\
&  +(\frac{1}{2}s+1)(\frac{1}{2}t+1)(\zeta_{1}\cdot k_{3}\zeta_{2}\cdot
k_{4}),\nonumber
\end{align}%
\begin{align}
K(\zeta_{1},k_{1};\zeta_{2},k_{2};\zeta_{3},k_{3};k_{4})  &  =(\frac{1}%
{2}s+1)(\frac{1}{2}t+1)\frac{1}{2}t(\zeta_{1}\cdot k_{3}\zeta_{2}\cdot
k_{1}\zeta_{3}\cdot k_{1}-\zeta_{2}\cdot k_{1}\zeta_{1}\cdot\zeta_{3}%
)\tag{46}\\
&  +(\frac{1}{2}s+1)(\frac{1}{2}t+1)\frac{1}{2}s(\zeta_{2}\cdot k_{3}\zeta
_{1}\cdot\zeta_{3}-\zeta_{1}\cdot k_{3}\zeta_{2}\cdot k_{3}\zeta_{3}\cdot
k_{1})\nonumber\\
&  +(\frac{1}{2}s+1)(\frac{1}{2}t+1)(\frac{1}{2}u+1)[\zeta_{1}\cdot k_{4}%
\zeta_{2}\cdot k_{1}\zeta_{3}\cdot k_{2}\nonumber\\
&  -\zeta_{1}\cdot k_{2}\zeta_{2}\cdot k_{3}\zeta_{3}\cdot k_{4}-\zeta
_{2}\cdot\zeta_{3}\zeta_{1}\cdot k_{2}+\zeta_{1}\cdot\zeta_{2}\zeta_{3}\cdot
k_{2}]\nonumber\\
&  -(\frac{1}{2}t+1)(\frac{1}{2}u+1)\frac{1}{2}t(\zeta_{1}\cdot\zeta_{2}%
\zeta_{3}\cdot k_{4}-\zeta_{1}\cdot k_{2}\zeta_{2}\cdot k_{1}\zeta_{2}\cdot
k_{4})\nonumber\\
&  -(\frac{1}{2}u+1)(\frac{1}{2}s+1)\frac{1}{2}s(\zeta_{1}\cdot k_{4}\zeta
_{2}\cdot k_{3}\zeta_{3}\cdot k_{2}-\zeta_{2}\cdot\zeta_{3}\zeta_{1}\cdot
k_{4}),\nonumber
\end{align}

\begin{align}
K(\zeta_{1},k_{1};\zeta_{2},k_{2};\zeta_{3},k_{3};\zeta_{4},k_{4})  &
=(\frac{1}{2}s+1)(\frac{1}{2}t+1)(\frac{1}{2}u+1)[-K^{(ss)}+\tag{47}\\
&  \frac{1}{2}s\{\zeta_{1}\cdot k_{3}\zeta_{2}\cdot k_{3}(\zeta_{3}\cdot
k_{1}\zeta_{4}\cdot k_{1}+\zeta_{3}\cdot k_{2}\zeta_{4}\cdot k_{2}%
)+\nonumber\\
&  \frac{1}{3}(\zeta_{1}\cdot k_{2}\zeta_{2}\cdot k_{3}\zeta_{3}\cdot
k_{1}-\zeta_{1}\cdot k_{3}\zeta_{2}\cdot k_{1}\zeta_{3}\cdot k_{2})(\zeta
_{4}\cdot k_{1}-\zeta_{4}\cdot k_{2})\}\nonumber\\
&  +\frac{1}{2}t\{\zeta_{2}\cdot k_{1}\zeta_{3}\cdot k_{1}(\zeta_{1}\cdot
k_{3}\zeta_{4}\cdot k_{3}+\zeta_{1}\cdot k_{2}\zeta_{4}\cdot k_{2}%
)+\nonumber\\
&  \frac{1}{3}(\zeta_{1}\cdot k_{3}\zeta_{2}\cdot k_{1}\zeta_{3}\cdot
k_{2}-\zeta_{1}\cdot k_{2}\zeta_{2}\cdot k_{3}\zeta_{3}\cdot k_{1})(\zeta
_{4}\cdot k_{3}-\zeta_{4}\cdot k_{2})\}\nonumber\\
&  +\frac{1}{2}u\{\zeta_{1}\cdot k_{2}\zeta_{3}\cdot k_{2}(\zeta_{2}\cdot
k_{1}\zeta_{4}\cdot k_{1}+\zeta_{2}\cdot k_{3}\zeta_{4}\cdot k_{3}%
)+\nonumber\\
&  \frac{1}{3}(\zeta_{1}\cdot k_{2}\zeta_{2}\cdot k_{3}\zeta_{3}\cdot
k_{1}-\zeta_{1}\cdot k_{3}\zeta_{2}\cdot k_{1}\zeta_{3}\cdot k_{2})(\zeta
_{4}\cdot k_{3}-\zeta_{4}\cdot k_{1})\}\nonumber\\
&  +\frac{st}{4}\frac{1}{\frac{u}{2}+1}(\zeta_{1}\cdot\zeta_{3}-\zeta_{1}\cdot
k_{3}\zeta_{3}\cdot k_{1})(\zeta_{2}\cdot\zeta_{4}-\zeta_{2}\cdot k_{4}%
\zeta_{4}\cdot k_{2})\nonumber\\
&  +\frac{st}{4}\frac{1}{\frac{u}{2}+1}(\zeta_{1}\cdot\zeta_{3}-\zeta_{1}\cdot
k_{3}\zeta_{3}\cdot k_{1})(\zeta_{2}\cdot\zeta_{4}-\zeta_{2}\cdot k_{4}%
\zeta_{4}\cdot k_{2})\nonumber\\
&  +\frac{su}{4}\frac{1}{\frac{t}{2}+1}(\zeta_{1}\cdot\zeta_{4}-\zeta_{1}\cdot
k_{4}\zeta_{4}\cdot k_{1})(\zeta_{2}\cdot\zeta_{3}-\zeta_{2}\cdot k_{3}%
\zeta_{3}\cdot k_{2})-\nonumber\\
&  \frac{st}{4}(\zeta_{1}\cdot\zeta_{3})(\zeta_{2}\cdot\zeta_{4})-\frac{tu}%
{4}(\zeta_{1}\cdot\zeta_{2})(\zeta_{3}\cdot\zeta_{4})\nonumber\\
&  -\frac{su}{4}(\zeta_{1}\cdot\zeta_{4})(\zeta_{2}\cdot\zeta_{3})],\nonumber
\end{align}
where $K^{(ss)}$ is the same kinematic factor that enters in the type I
superstring \ and can be found in \cite{15}. It is easy to show that
$K^{(ss)}$ is suppressed in energy in the high energy expansion. Finally, the
4-point function of a spin-two, a vector and two tachyons is calculated to be%
\begin{align}
\mathcal{T}_{tensor}  &  =%
{\textstyle\int}
{\textstyle\prod_{i=1}^{4}}
dx_{i}<\zeta_{\mu\nu}\partial X^{\mu}\partial X^{\nu}e^{ik_{1}X}\zeta_{\nu
}\partial X^{\nu}e^{ik_{2}X}e^{ik_{3}X}e^{ik_{4}X}>\tag{48}\\
&  =\frac{\Gamma(-\frac{s}{2}-1)\Gamma(-\frac{t}{2}-1)}{\Gamma(\frac{u}{2}%
+2)}[2(\frac{1}{2}t+1)\frac{1}{2}u(\frac{1}{2}u+1)(\zeta_{\mu\nu}k_{2}^{\mu
}\zeta^{\nu})\nonumber\\
&  -(\frac{1}{2}t+1)\frac{1}{2}u(\frac{1}{2}u+1)(\zeta_{\mu\nu}k_{2}^{\mu
}k_{2}^{\nu})(\zeta\cdot k_{1})+(\frac{1}{2}s+1)\frac{1}{2}u(\frac{1}%
{2}u+1)(\zeta_{\mu\nu}k_{2}^{\mu}k_{2}^{\nu})(\zeta\cdot k_{3})\nonumber\\
&  -2(\frac{1}{2}s+1)(\frac{1}{2}t+1)(\frac{1}{2}u+1)(\zeta_{\mu\nu}k_{3}%
^{\mu}\zeta^{\nu})+2(\frac{1}{2}s+1)(\frac{1}{2}t+1)(\frac{1}{2}%
u+1)(\zeta_{\mu\nu}k_{2}^{\mu}k_{3}^{\nu})(\zeta\cdot k_{1})\nonumber\\
&  -2(\frac{1}{2}s)(\frac{1}{2}s+1)(\frac{1}{2}u+1)(\zeta_{\mu\nu}k_{2}^{\mu
}k_{3}^{\nu})(\zeta\cdot k_{3})-(\frac{1}{2}s)(\frac{1}{2}s+1)(\frac{1}%
{2}t+1)(\zeta_{\mu\nu}k_{3}^{\mu}k_{3}^{\nu})(\zeta\cdot k_{1})\nonumber\\
&  +(\frac{1}{2}s)(\frac{1}{2}s+1)(\frac{1}{2}s-1)(\zeta_{\mu\nu}k_{3}^{\mu
}k_{3}^{\nu})(\zeta\cdot k_{3}).\nonumber
\end{align}
By using $s=4E^{2}$, $t\simeq-4\sin^{2}\frac{\phi_{CM}}{2}E^{2}$ and
$u\simeq-4\cos^{2}\frac{\phi_{CM}}{2}E^{2}$ in the high energy limit, the
transverse components of the high energy limits of eqs.(45), (46), (47) and
(48) are calculated to be $[-2E^{3}\sin\phi_{CM}]^{1+1}\mathcal{T}(2)$,
$[-2E^{3}\sin\phi_{CM}]^{1+1+1}\mathcal{T}(3)$, $[-2E^{3}\sin\phi
_{CM}]^{1+1+1+1}\mathcal{T}(4)$ and $[-2E^{3}\sin\phi_{CM}]^{2+1}%
\mathcal{T}(3).$ These are remarkably consistent with the prediction of eq.(43).

The second master formula for high energy scattering is valid for processes
with vertices not containing $\partial X^{L}$, or equivalently, for those
amplitudes whose true leading order in energy are the same as the naive
leading order in energy. The high energy string-tree scattering amplitudes of
this type with one tensor $v_{2}$ and three tachyons $v_{1,3,4}$ are
calculated to be (we list amplitudes for the $s-t$ channel only)%
\begin{equation}
\mathcal{T}_{n}^{sub}=\prod_{a}\zeta_{a}\cdot([-\frac{t}{2}]^{l_{a}}%
k_{1}+[\frac{s}{2}]^{l_{a}}k_{3})\mathcal{T}(n), \tag{49}%
\end{equation}
where $\zeta_{a}$, which is either $e^{T}$ or $e^{L}$, corresponds to the
polarization of $\partial^{l_{a}}X$ in the vertex operator $v_{2}$ at mass
level $m^{2}=2(n-1)$, $n=\Sigma l_{a}$ . For example, $\mathcal{K}_{(LT)}^{3}$
, which contains a vertex $\partial X$ $^{(T}\partial^{2}X$ $^{L)}$, can be
rewritten as%
\begin{equation}
\mathcal{K}_{(LT)}^{3}=(-\frac{t}{2}k_{1}+\frac{s}{2}k_{3})\cdot
e_{(T}([-\frac{t}{2}]^{2}k_{1}+[\frac{s}{2}]^{2}k_{3})\cdot e_{L)}=E^{9}%
\sin^{3}\phi_{CM}, \tag{50}%
\end{equation}
which is correctly predicted by eq.(43). It is straightforward to write down a
general formula for the four-tensor scattering amplitudes of these types at
arbitrary mass levels. Note that eq.(49) includes processes that are not the
leading high energy scattering amplitudes at each fixed mass level considered
in eq.(43). For example, eq.(49) gives%
\begin{equation}
\mathcal{K}_{(TT)}^{2}=(-\frac{t}{2}k_{1}+\frac{s}{2}k_{3})\cdot e_{T}%
([-\frac{t}{2}]^{2}k_{1}+[\frac{s}{2}]^{2}k_{3})\cdot e_{T}=8E^{8}\sin^{2}%
\phi_{CM}, \tag{51}%
\end{equation}
which is not given by eq.(43). This scattering amplitude is of subleading
order in energy at the mass level $m^{2}=4$. Note that the superscripts of
$\ \mathcal{K}_{(LT)}^{3}$ and $\mathcal{K}_{(TT)}^{2}$ represent the naive
(or true) leading orders of the scattering amplitudes defined in the paragraph
after eq.(14).

\section{Conclusion}

We have shown that saddle point calculations of high energy string scattering
amplitudes of eqs.(21)-(24) \cite{2,3,11} are not consistent with the
zero-norm state calculations of high energy stringy Ward identities of
eqs.(8)-(11) \cite{7,8}. In this paper, we have also given the correct saddle
point calculation, eqs.(34)-(40) which are consistent with our previous
calculation based on a different method \cite{7,8}, to illustrate the
importance of subleading energy power factor in front of the exponential of
the high energy string amplitude. This power factor is crucial to recover the
stringy Ward identities and the linear relations among scattering amplitudes
of different string states conjectured by Gross \cite{3}.

Based on the tree-level stringy Ward identities derived from the decoupling of
two types of zero-norm states, it was conjectured \cite{7,8} that there is
only one independent component of high energy scattering amplitude at each
fixed mass level. All other components of high energy scattering amplitudes
are proportional to it. This conjecture was explicitly proved for the mass
levels $m^{2}=4,6$ \cite{7,8}. If this conjecture is valid to all higher mass
levels, our master formula, eq.(43), determines all high energy string-tree
scattering amplitudes in terms of those of tachyons- another conjecture by
Gross \cite{3}. It is worth noting that if all stringy propagating modes
contribute at least one high energy scattering amplitude, Then eq.(43) applies
to all particles in the string spectrum.

While the importance of zero-norm states in string theory has been largely
underestimated, we expect that a clearer understanding of zero-norm states
will help us to uncover the fundamental symmetry of string theory.

\section{Acknowledgments}

This work is supported in part by the National Science Council, Taiwan, R.O.C.
We would like to thank Hiroyuki Hata, Hsien-chung Kao, and Yutaka Matsuo for
discussions. CT and JC would also like to thank NCTS/TPE for the hospitality.


\begin{thebibliography}{99}
\bibitem {13}Another correction one has to take into account is the saddle
point $\widehat{\lambda}\approx-\frac{t}{s}\approx\sin^{2}\frac{\phi_{CM}}{2}$
in eq.(21) identified by authors of Refs \cite{2,3,11}. The saddle point we
identified in eq.(35) is, however, $x_{0}=\frac{1}{1-\tau}\approx\frac{1}%
{\cos^{2}\frac{\phi_{CM}}{2}}$. Since our convention of Mandelstam variables
is different from Refs \cite{2,3,11} by intercganging $t\longleftrightarrow
u$, we believe that the correct saddle point of Refs \cite{2,3,11} should be
$\widehat{\lambda}_{corr}\approx\frac{1}{\sin^{2}\frac{\phi_{CM}}{2}}$.
\end{thebibliography}
\end{document}